# Laboratory Trajectories Improve Kidney Failure Risk Estimation

Morgan Sanchez, M.A.[1,2], James A. Diao, M.D., M.Phil.[1-3], Jesse Cummings, M.Eng.[2], Maya Makov-Assif, M.D.[4], Liat Antwarg Friedman, Ph.D.[1,2,4], Seffi Cohen, Ph.D.[1,2,4], Aashna P. Shah, M.S.[1,2,5], Ben Reis, Ph.D.[1,2,6], Ran D. Balicer, M.D., M.P.H.[1,4,7] *, Noa Dagan, M.D., Ph.D.[1,4,8] *, and Arjun K. Manrai, Ph.D.[1,2] *

[1] Ivan and Francesca Berkowitz Family Living Laboratory, Harvard Medical School and Clalit Research Institute, Boston, MA, USA and Tel Aviv, Israel

[2] Department of Biomedical Informatics, Harvard Medical School, Boston, MA, USA

[3] Department of Medicine, Brigham and Women's Hospital, Boston, MA, USA

[4] Clalit Research Institute, Innovation Division, Clalit Health Services, Tel Aviv, Israel

[5] Department of Systems Biology, Harvard Medical School, Boston, MA

[6] Predictive Medicine Group, Boston Children's Hospital, Boston, MA, USA

[7] School of Public Health, Faculty of Health Sciences, Ben Gurion University of the Negev, Be'er Sheva, Israel

[8] Faculty of Computer and Information Science, Ben Gurion University, Be'er Sheva, Israel

* Co-senior authors

Correspondence:
Arjun K. Manrai, Ph.D.
Department of Biomedical Informatics, Harvard Medical School
10 Shattuck St., Boston, MA, 02115
Arjun_Manrai@hms.harvard.edu

## Abstract

Accurate kidney failure risk assessment is critical to timely intervention in chronic kidney disease (CKD). Existing equations (e.g. Kidney Failure Risk Equation; KFRE) rely on single laboratory measurements to estimate short- and long-term kidney failure risk, leaving longitudinal laboratory patterns unused. Here we introduce Clalit Longitudinal Assessment of Risk of Kidney Failure (CLARK), an interpretable longitudinal extension of latest-value methods which incorporates routinely collected repeat laboratory measures. We develop CLARK using data from 5.4 million individuals, identifying 270,009 patients with CKD to create one of the largest longitudinal CKD cohorts to date, with 12,087 kidney replacement therapy initiation events and a median follow-up of 10.4 years. Across laboratory configurations and prediction horizons, CLARK demonstrated improved discrimination over static models (e.g., 2-year average precision 0.541 vs 0.516 in the eGFR-only setting). At intervention thresholds, trajectory-based models improved identification of high-risk patients, especially for longer-term prediction, suggesting that interpretable longitudinal laboratory features may enhance kidney failure risk assessment through improved identification of patients most likely to benefit from timely intervention.

## Introduction

Chronic kidney disease (CKD) affects over 700 million or 1 in 11 individuals globally[1]. CKD is characterized by progressive kidney function decline that may lead to kidney failure requiring dialysis or kidney transplantation. Although numerous interventions are available to slow CKD progression, such as blood pressure control, renin-angiotensin system blockade, and dietary modification, as well as to prepare for initiation of dialysis or transplant, these interventions can be costly and may pose risks to the patient[2–4]. Consequently, to identify patients at high risk of progression and determine when to recommend an intervention, clinical guidelines recommend validated equations for estimating short-term and long-term risk of kidney failure[5,6].

The Kidney Failure Risk Equation (KFRE), featured in both the Kidney Disease: Improving Global Outcomes (KDIGO) and National Institute for Health and Care Excellence guidelines, is the predominant equation used by clinicians to estimate 2- and 5-year risk of kidney replacement therapy (KRT) in patients with CKD stages 3–5[5,6]. Several KFRE variants were developed[7], with the best-performing equations subsequently validated in international populations[8]. These validated models use age, sex, and a limited set of laboratory values measured at a single time point. However, they neglect information contained in repeated laboratory measurements, which are routinely collected for many patients with or at risk of kidney disease.

A growing body of work suggests that repeated laboratory measurements contain prognostic information beyond that captured by single measurements both for patients with kidney disease and more broadly[9–13]. For example, eGFR slope and latest eGFR have been shown to be independently associated with kidney failure[9], and changes in urinary albumin-creatinine ratio (UACR) have also been linked to kidney outcomes[11]. More broadly, repeated laboratory values can be used to define patient-specific "setpoints" associated with common diseases and comorbidities, including CKD[13].

For kidney failure risk estimation in particular, prior studies have incorporated longitudinal laboratory data by updating predictions using the latest available measurements[14] or by expanding beyond the input variables used in KFRE[15]. However, latest-value updating remains fundamentally snapshot-based, and studies using expanded longitudinal inputs do not isolate whether performance gains arise from repeated measurements themselves, expanded input features, or more advanced modeling methods. Further, prior longitudinal studies utilize small to mid-sized cohorts ranging from 3,000 to 80,000 individuals.

In the present study, we introduce CLARK (Clalit Longitudinal Assessment of Risk of Kidney Failure), a longitudinal extension of latest-value risk models such as KFRE, developed and evaluated on a cohort of over 270,000 individuals with CKD. We extend the existing static equations in a controlled manner by augmenting them with interpretable features derived from repeated laboratory measurements and their trajectories. Specifically, we construct features that capture patient-specific baseline and recent levels, rates of change, and deviation from patient-specific expected trajectories. These approaches build directly on established concepts such as eGFR slope and latest value, while remaining computationally simple and clinically interpretable. We demonstrate the added predictive value and clinical impact of incorporating trajectory-based features for both short- and long-term KRT risk estimation through carefully controlled comparison not just to validated KFRE equations used in current clinical practice, but also to parallel static equations developed in the same population and with the same modeling approach as CLARK. This design allows us to directly assess the contribution of longitudinal trajectory information and identify where performance improvements arise.

# Results

### *Population Demographics*

The study population was derived from the electronic health records of 5.4 million members of Clalit Health Services (CHS), the largest health management organization in Israel, according to the criteria outlined in Figure 1. The final study population, described in Table 1, included 270,009 adults with CKD of any stage and no prior history of KRT.

In the overall study population, participants were distributed across CKD stages, with a substantial proportion in earlier stages (stages 1–2: 63.2%) and considerable representation in moderate stages (stage 3a: 23.2%, stage 3b: 9.6%), while fewer individuals were classified as stage 4 (3.3%) or stage 5 (0.6%). Median age was 72 years, and 53.6% were female. Median serum creatinine (SCr) and UACR at index were 0.95 mg/dL (IQR 0.73–1.22) and 38 mg/g (IQR 14.37–103.97), respectively. Rates of missingness among single, latest lab values were low (0%–3.5%), with the exception of UACR, which was missing in 7% of participants, and bicarbonate, which was missing in 84% and therefore excluded from further analysis.

During the historical observational period prior to the index date, repeated measurements for eGFR were dense and extensive for the vast majority of participants. Overall, 97% of participants had eGFR values spanning at least 5 years, and 88% had more than 10 measurements. Calcium, phosphate, and albumin measurements also had substantial temporal coverage (>5 years: 89%–92%), although they were less densely sampled (>10 measurements: 53.5%–61.6%). UACR measurements also showed substantial coverage, but were less frequently obtained and spanned shorter durations (>5 years: 66%; >10 measurements: 20%). Figure 2a further characterizes measurement duration and frequency during this historical observational period.

Following the index date, participants were observed during a median follow-up period of 10.4 years (IQR 4.7–10.4). During this period, 12,087 participants (4.5%) initiated KRT, including 2,695 (1.0%) within 2 years and 6,251 (2.3%) within 5 years of the index date.

Among individuals who initiated KRT during follow-up, demographic characteristics at index were broadly similar to the overall population, although several differences were observed. The proportion of female participants was lower (39.5%), consistent with prior KRT populations[16], and the median age was lower (68 years), with a greater concentration in the 45–64 and 65–74 age groups and fewer individuals aged 75 years or older. Baseline laboratory values reflected more advanced kidney disease, with higher median SCr (1.51 mg/dL) and UACR (304.10 mg/g). CKD stage distribution was correspondingly shifted toward more advanced disease among individuals who initiated KRT, with a greater proportion in stages 3b–5 (stage 3b: 22.8%, stage 4: 20.6%, stage 5: 7.1%) and a smaller, though still substantial, proportion in earlier stages (stages 1–2: 30.3%). Missingness of laboratory values was lower in this group, consistent with more intensive monitoring among patients who ultimately progress to kidney failure.

### *Estimated Glomerular Filtration Rate Trajectories*

Given the dense longitudinal eGFR measurements in this population, we first examined population-level patterns of kidney function over time prior to the index date (Figure 2b). Participants were stratified by their most recent eGFR value (<30, 30–44, 45–59, ≥60 ml/min/1.73 $m^2$), enabling comparison of individuals with similar current kidney function, who are treated similarly in latest-value-based risk estimation approaches. Within each stratum, median eGFR at index was, by design, similar between participants who progressed to KRT and those who did not. However, their preceding trajectories diverged substantially.

At the population level, individuals who ultimately progressed to KRT exhibited sustained and often accelerating declines in eGFR over several years preceding the index date, whereas non-progressors showed relatively stable or slowly declining kidney function. These differences emerged several years prior to index and persisted across all eGFR strata, indicating that patients with similar current kidney function may follow markedly different disease trajectories.

This observation highlights a key limitation of latest-value approaches, in which a single measurement compresses heterogeneous longitudinal patterns into a static snapshot. Capturing these dynamics requires features that summarize baseline kidney function, rates of decline, variability, and changes in trajectory over time, which are not represented by the most recent eGFR alone. These results were consistent with prior descriptions of heterogeneous and nonlinear CKD progression[17] and informed the construction of trajectory-based features derived from longitudinal eGFR measurements as well as the remaining KFRE laboratory inputs (Figure 2c).

### *Baseline and Experimental Models*

To assess whether incorporating these trajectory-based features improves kidney failure risk estimation, we compared trajectory-based and latest-value models. Model performance was evaluated using discrimination, calibration, and net reclassification improvement (NRI) at guideline-recommended and quantile-based intervention thresholds across several lab sets, largely mirroring KFRE variants.

The published KFREs are externally validated, widely adopted clinical equations. However, since KFRE models were developed in a different patient population using a different modeling approach, differences in data distribution and methodology could confound direct comparisons between the latest-value approach and the trajectory-based approach. Therefore, to establish latest-value baselines and control for population-specific effects and modeling choices, we evaluated two families of latest-value risk estimation models which rely on single measurements of laboratory values: published KFRE equations and Clalit Static Assessment of Risk of Kidney Failure (CSARK) models. (Figure 2d)

The CSARK models use the same input features as KFRE (age, sex, and latest laboratory values), but are developed using the Clalit development cohort. We verified that CSARK models and KFRE models showed comparable discriminative performance and focused on comparisons between CLARK and CSARK to evaluate the added predictive power of trajectory-based features. This ensures that any observed performance differences reflect the added value of temporal features rather than artifacts of dataset composition, modeling choices, or optimization strategies.

### *Latest Value Models Achieve Moderate Discrimination and Calibration*

We evaluated two variants of KFRE and CSARK, each with input features mirroring published KFRE models: $KFRE_{eGFR\text{-}only}$, which uses age, sex, and latest eGFR, and $KFRE_{eGFR+UACR}$, which additionally incorporates latest UACR. Due to limited bicarbonate data availability, we were unable to evaluate the full 6-lab KFRE model or produce a 6-lab CSARK model. Instead, we trained and evaluated $CSARK_{5\text{-}lab}$, which incorporates eGFR and UACR as well as serum albumin, calcium, and phosphate. All models were evaluated at 2- and 5-year prediction horizons.

CSARK models showed moderate discrimination across all KRT prediction horizons and laboratory settings, with discrimination improving as the laboratory set expanded. Average precision (AP) increased with longer prediction horizons, whereas c-statistics declined as the horizon extended. Calibration improved with expansion of the laboratory set and with shorter prediction horizons and was consistently improved relative to KFRE.

Relative to KFRE, CSARK demonstrated comparable or improved discrimination (Table 2). For 2-year KRT prediction, KFRE achieved an average precision of 0.520 (95% CI: 0.516–0.525) in the eGFR-only setting and 0.536 (95% CI: 0.531–0.540) in the eGFR+UACR setting. In comparison, $\text{CSARK}_{\text{eGFR-only}}$ showed similar discrimination (AP 0.516; 95% CI: 0.511–0.521), whereas discrimination for $\text{CSARK}_{\text{eGFR+UACR}}$ (AP: 0.549; 95% CI: 0.545–0.553) was improved over KFRE. For predicting KRT within 5 years, KFRE achieved an average precision of 0.536 (95% CI: 0.533–0.539) in the eGFR-only setting and 0.580 (95% CI: 0.577–0.583) in the eGFR+UACR setting. CSARK again showed comparable discrimination in the eGFR-only setting (AP 0.533; 95% CI: 0.530–0.536) and improved discrimination in the eGFR+UACR setting (AP 0.594; 95% CI: 0.591–0.597).

In the 5-lab setting, CSARK achieved an average precision of 0.582 (95% CI: 0.577–0.586) for the 2-year horizon and an average precision of 0.617 (95% CI: 0.614–0.620). Full precision-recall curves for CSARK are shown in Figure 3d, and complete discrimination and calibration metrics are provided in Table 2.

### *Trajectory-Based Features Improve Discrimination and Calibration*

Compared with CSARK, CLARK models generally demonstrated superior discrimination and calibration across prediction horizons and laboratory settings. For both modeling approaches, expanding the laboratory feature set was associated with improved performance. The addition of UACR was associated with larger gains for 5-year prediction than for 2-year prediction, while improvements from further expansion to the 5-lab setting were more gradual for 2-year prediction (Table 2, Figure 3a).

For prediction of KRT within 2 years, average precision was 0.541 (95% CI: 0.537–0.545) for $\text{CLARK}_{\text{eGFR-only}}$, 0.564 (95% CI: 0.560–0.568) for $\text{CLARK}_{\text{eGFR+UACR}}$, and 0.585 (95% CI: 0.581–0.589) for the 5-lab model. Relative to CSARK, CLARK improved average precision by 0.024 (95% CI: 0.023–0.026), 0.015 (95% CI: 0.014–0.016), and 0.004 (95% CI: 0.002–0.005) for the eGFR-only, eGFR+UACR, and 5-lab models, respectively.

For prediction of KRT within 5 years, discrimination was higher overall. Average precision was 0.567 (95% CI: 0.564–0.570) for $\text{CLARK}_{\text{eGFR-only}}$, 0.615 (95% CI: 0.612–0.618) for $\text{CLARK}_{\text{eGFR+UACR}}$, and 0.629 (95% CI: 0.626–0.632) for the 5-lab model. Improvements relative to CSARK were 0.034 (95% CI: 0.033–0.035), 0.021 (95% CI: 0.020–0.021), and 0.012 (95% CI: 0.011–0.012) for the respective laboratory settings.

Across both prediction horizons, the largest gains from trajectory-based modeling were observed in the eGFR-only setting, with diminishing benefit as additional laboratory measurements were incorporated. Notably, the increase in performance from adding UACR was larger for CSARK than for CLARK, particularly for 5-year prediction. Calibration improved modestly with the inclusion of longitudinal features, as reflected by consistently lower or comparable Brier scores, and c-statistics were higher for CLARK in all comparisons. Full performance statistics are presented in Table 2, and average precision comparisons and precision-recall curves are shown in Figure 3.

### *Reclassification at Guideline-Based Interventional Risk Thresholds*

Although we observed improvements in threshold-agnostic performance metrics such as discrimination and calibration, these measures do not directly capture how predictions translate into clinical decisions.

KDIGO guidelines define clinically actionable risk thresholds that guide clinical interventions in patients with chronic kidney disease, including nephrology referral at ≥3% and ≥5% estimated 5-year risk of KRT, initiation of team-based care at ≥10% estimated 2-year risk, and planning for KRT at ≥40% estimated 2-year risk. To evaluate clinical impact, and specifically how CLARK alters patient assignment across these decision-relevant thresholds relative to CSARK, we compared models using net reclassification improvement across eGFR-only, eGFR+UACR, and 5-lab settings.

NRI quantifies the net change in correct risk classification between models at a given threshold. Given the low event rate and substantial imbalance between progressors and non-progressors, we evaluated reclassification separately for individuals who progressed to KRT and those who did not.

Across guideline-based thresholds, CLARK changed patient classification relative to CSARK, with effects differing between individuals who progressed to KRT and those who did not (Figure 4a). $NRI_{KRT}$, which reflects improved identification of individuals who progressed to KRT, was generally positive and most consistent at 5-year risk thresholds across laboratory settings. In contrast, $NRI_{No\ KRT}$, which captures changes among non-progressors, was frequently negative, indicating increased false positive classification, accompanying the improved identification of progressors.

Differences between prediction horizons were pronounced. At 5-year thresholds, $NRI_{KRT}$ improvements were observed across laboratory configurations, including the 5-lab setting, suggesting that trajectory-based features provide consistent benefit for longer-term risk stratification. In contrast, at 2-year thresholds, $NRI_{KRT}$ gains were smaller and less consistent, with minimal or negative effects at the highest-risk threshold for KRT planning (≥40% 2-year risk).

Across laboratory configurations, $NRI_{KRT}$ gains were largest in the eGFR-only setting and attenuated with the addition of UACR and other laboratory features, while $NRI_{No\ KRT}$ remained near zero or slightly negative. Together, these findings suggest that trajectory-based features primarily improve identification of individuals at longer-term risk of progression, but may increase false positive classification among non-progressors, particularly at higher-risk thresholds when KDIGO risk cutoffs are used.

### *Reclassification at Quantile-Based Interventional Risk Thresholds*

While guideline-based thresholds provide clinically interpretable decision points, they are defined using KFRE-derived risk estimates and therefore depend on the properties of a specific model. In practice, intervention decisions may also reflect resource constraints, outcome prevalence, and target intervention scales. To evaluate the impact of trajectory-based modeling under a model-agnostic framework and across a range of intervention scales, we assessed reclassification using quantile-based strategies, where interventions are applied to the top k% of individuals based on predicted risk (Figure 4b).

Similar to guideline-based analyses, CLARK generally improved identification of individuals who progressed to KRT across both 2-year and 5-year prediction horizons. $NRI_{KRT}$ was positive across most quantile thresholds and laboratory configurations. However, unlike guideline-based reclassification, where improvements in $NRI_{KRT}$ were sometimes offset by increased false positives, $NRI_{No\ KRT}$ was uniformly small and not distinguishable from zero across all quantiles. As a result, reclassification under quantile-based strategies was driven almost entirely by improved identification of progressors, without meaningful changes in classification of non-progressors.

The magnitude and stability of improvements varied by prediction horizon and intervention scale. For 5-year prediction, $NRI_{KRT}$ gains were consistently positive across quantiles and laboratory settings, attenuating modestly with expanded laboratory sets but remaining present even in the 5-lab configuration. For interventions on the top 10% of participants, improved identification of KRT progressors was similar regardless of laboratory setting, reaching 1.8%, corresponding to 18 additional KRT cases identified per 1000 true KRT cases missed by the static model. This indicates robust improvements in long-term risk stratification. In contrast, 2-year results were smaller and more variable, with limited improvement at lower quantiles and inconsistent effects at higher intervention thresholds.

Because quantile-based strategies directly correspond to intervention capacity, these results suggest that trajectory-based models primarily improve identification of individuals at longer-term risk of progression, particularly when interventions are applied at moderate to larger scales. The absence of meaningful changes in $NRI_{No\ KRT}$ indicates that these gains are achieved without increasing false positive classification among non-progressors.

### *Interpretable Sources of Model Improvement*

To better understand the source of these performance gains, we decomposed improvements in average precision into sequential feature enrichment steps, beginning with the base latest-value models and incrementally adding groups of correlated trajectory-based features representing related clinical concepts (Figure 4c).

In the first step, we incorporated the trajectory-derived recent eGFR group. This step represents improvement due to more robust estimation of current kidney function, derived from longitudinal measurements rather than a single latest value as used in CSARK. This group produced substantial gains in both 2- and 5-year prediction, particularly in the eGFR-only setting.

Further improvements were driven by feature groups capturing baseline kidney function and variability in eGFR over time, both concepts not available to single timepoint models such as CSARK or KFRE. Baseline eGFR features, which reflect long-term and pre-decline kidney function, contributed most strongly to 5-year prediction across all laboratory settings. Measures of eGFR variability also improved performance for both prediction horizons, with the largest effects observed in the eGFR-only and eGFR+UACR settings.

In contrast, expansion to the full set of trajectory-based features resulted in relatively modest incremental gains. This indicates that a large portion of CLARK's predictive advantage can be attained using a small set of clinically interpretable features: a more stable estimate of current kidney function and trajectory-derived measures of baseline function and variability.

## Discussion

In a large, longitudinal cohort of over 270 thousand CKD patients with more than a decade of follow-up, laboratory histories carried substantial prognostic information beyond single latest measurements. We developed CLARK, an interpretable longitudinal extension of latest-value kidney failure risk models, by augmenting existing risk-estimation approaches with simple trajectory-based features derived from routinely collected laboratory inputs. CLARK consistently improved discrimination and calibration across prediction horizons and laboratory configurations, indicating that trajectory-based features capture prognostic information across increasingly comprehensive sets of input laboratories.

The potential clinical relevance of these improvements is supported by reclassification analyses at both guideline-based and quantile-based intervention thresholds. At KDIGO-aligned thresholds, trajectory-based modeling improved identification of individuals who progressed to KRT, although these gains were sometimes offset by increased false positive classification among non-progressors. In contrast, under quantile-based intervention approaches, which more directly reflect clinical resource constraints, improvements were driven almost entirely by improved identification of progressors, without measurable increases in false positives. These gains were most consistent for 5-year prediction, suggesting that trajectory-based features are particularly informative for longer-term risk stratification.

The magnitude of improvement varied across laboratory configurations. Gains were largest in the eGFR-only setting and attenuated with the addition of UACR and other laboratory measurements. This pattern suggests that trajectory-derived features capture a substantial portion of the prognostic information provided by albuminuria and other biomarkers, while still contributing incremental signal when richer data are available. This highlights an important practical implication, that longitudinal modeling may be especially valuable in settings where key biomarkers such as UACR are unavailable or infrequently measured.

Decomposition of model performance revealed that a small set of trajectory-derived features captured most of the predictive gain. Building on prior work showing that eGFR slope provides prognostic information beyond the latest eGFR[9], we extended this approach by incorporating a broader set of trajectory-derived laboratory features. Among these features, measures of eGFR variability contributed the largest gains, highlighting fluctuations and instability in kidney function as a key source of prognostic signal. Features capturing recent kidney function provided more stable estimates of current disease state, suggesting that single latest values may be affected by measurement noise. In addition, features reflecting baseline kidney function contributed most strongly to longer-term prediction, consistent with the role of long-term functional reserve in disease progression. Although eGFR slope was also examined, variability and baseline measures contributed more strongly to prediction, suggesting that these aspects of kidney function capture clinically relevant dynamics not reflected in average rates of decline.

Several aspects of the study design support these interpretations. First, CLARK was developed as a controlled longitudinal extension of latest-value risk models, preserving interpretability and alignment with guideline-based inputs. We compared trajectory-based models against parallel latest-value models developed in the same population, using the same prediction horizons, laboratory configurations, and modeling approach. Unlike prior studies, which expand input data or use complex architectures[15], this design allowed us to distinguish the added value of longitudinal trajectory information from gains attributable to expanded predictors, different cohorts, or more flexible modeling methods. Second, sequential feature enrichment experiments enabled us to examine which longitudinal concepts contributed most to prediction, rather than treating longitudinal modeling as a single black-box extension. Third, the scale and richness of the Clalit CKD cohort enabled robust evaluation of trajectory-based kidney failure risk assessment. Compared with other large CKD cohorts such as the VA CKD cohort, which is predominantly male and contains substantial missingness, the Clalit CKD cohort comprises a more balanced sex representation and more complete laboratory histories.

In addition to its strengths, this study has several limitations. First, although the Clalit cohort is large and multi-ethnic, it represents a single healthcare system, and external validation is required to assess generalizability to other populations, care settings, laboratory measurement practices, and healthcare delivery systems. Second, because this was an observational study using routinely collected electronic health record data rather than a prospective study with standardized follow-up, laboratory measurements reflect clinical care patterns and healthcare utilization. For example, patients with more frequent testing may differ systematically from those with sparse laboratory histories. Further, the magnitude of benefit from trajectory-based features may differ in settings with less frequent laboratory measurement or greater missingness. Third, outcomes are derived from available data within the Clalit system, meaning KRT events occurring outside the available data may be missed or delayed. Finally, bicarbonate measurements were infrequent in this population and could not be evaluated, limiting comparison with the full 6-variable KFRE.

In this study, simple, interpretable features derived from trajectories of routinely collected laboratory measurements consistently improved kidney failure risk estimation beyond latest-value models. These gains were driven primarily by improved identification of individuals at risk of progression and were most pronounced in settings lacking albuminuria, while remaining present across more comprehensive laboratory configurations. Incorporating trajectory-based features provides a direct extension of existing risk equations, capturing temporal patterns that clinicians already recognize while preserving interpretability and alignment with established clinical frameworks.

# Online Methods

***Setting***

We leveraged a longitudinal clinical research dataset collected, stored, and processed by Clalit Health Services, which contains over 20 years of anonymized demographic data, diagnostic and procedure codes, laboratory measurements, visit records, and membership dates for over half of the Israeli population, totaling over 5.4 million patients. CHS is both a payer and provider of clinical services and operates thousands of outpatient and inpatient clinical sites throughout Israel. The study was performed with the approval of CHS's Institutional Review Board.

***Study Population***

From the CHS dataset, we extracted a primary study population of adults with evidence of decreased kidney function and no history of KRT. Specifically, using a reference date of January 1, 2014, we extracted all participants with at least one SCr measurement within the three years before the reference date, and the date of each participant's last SCr measurement before the reference date was used as their index date. For participant exclusion and index date determination, SCr measurements taken during pregnancy, hospitalization, or when the participant was outside the ages of 25 to 90 were excluded. As a result, the population was restricted to ages 25 to 90 at index. We further excluded participants missing demographic data, participants with less than 1 year of continuous membership before their index date, and participants with a recorded kidney transplant or dialysis procedure or diagnostic code prior to their index date.

Finally, we extracted patients with CKD of any stage at index. To obtain CKD diagnoses, we calculated eGFR via the 2021 CKD-EPI equation[18] for all SCr values and extracted all UACR measurements prior to the index date, excluding measurements taken during childhood (age < 18), pregnancy, and hospitalizations for both SCr and UACR. CKD was defined as two or more consecutive measures spanning at least three months of eGFR <60 ml/min/1.73 m$^2$ or UACR ≥30 mg/g. Figure 1 visualizes the primary components of participant filtering.

### *Outcome Definitions*

For this study, we modeled kidney disease progression using survival analysis, where the event of interest is the initiation of KRT (i.e. first kidney transplant or dialysis treatment after the index date). To determine KRT initiation labels, we extracted all diagnostic and procedure codes related to KRT and their associated dates for all participants and kept only the earliest date for each participant. We also extracted loss of followup, which is defined as any lapse in CHS membership prior to the event of interest, and death. For each participant, we recorded only the first event, either KRT, loss of followup, or death as event status (KRT or non-KRT) and event time (i.e the time elapsed between the index date and the event). The end of study date was defined as the date of the last recorded KRT event of any participant. Participants that did not experience any event were censored at the end of the study.

In addition to extracting event time and status, we also derived several binary outcomes: whether the participant experienced KRT within 2 and 5 years of their index date.

### *Feature Extraction*

We extracted three types of features: demographic features, latest value features, and trajectory-based features. Demographic features, which are included as features in all models, include sex and age at index.

#### *Latest Value Features*

Latest value features are single lab value measurements. Specifically, we extracted the last values on or before the index date for eGFR, UACR, albumin, calcium, phosphate, and bicarbonate, all labs used in KFRE equations. Measurements taken during childhood, pregnancy, and hospitalizations were excluded for all labs in extracting latest measurements. Bicarbonate was excluded from further analysis due to lack of measurement in the CHS population.

#### *Trajectory-Based Features*

Trajectory-based features are defined as features extracted from a participant's full or partial lab history, excluding measurements taken during childhood and pregnancy. Unlike the latest value features, however, measurements taken during hospitalizations were not excluded.

For each laboratory type (e.g. eGFR, UACR, albumin, etc.), we computed several features including measures of central tendency, variability, extrema, and both linear and nonlinear fitted trends. We extracted basic statistics such as minimum, maximum, mean, median, and standard deviation values over all time prior to index and within 3 and 5 years prior to the index date. We fit linear models over all time and extracted fit parameters such as slope, intercept, and mean squared error. We also detected "changepoints" using binary segmentation of the trajectories and extracted basic statistical measures before and after detected changepoints. We fit linear models to lab trajectories before and after changepoints and extracted slopes and intercepts from these lines. Finally, we fit exponential curves to the full lab trajectories and extracted fit parameters as well as slope, acceleration, and intercept at the index date. A visualization of trajectory-based feature extraction can be found in Figure 2c.

### *Model Comparison*

To determine whether trajectory-based features provide incremental predictive value beyond static risk models, we evaluated three laboratory configurations: (1) eGFR-only, (2) eGFR and UACR, and (3) a 5-lab set including eGFR, UACR, albumin, calcium, and phosphate. These configurations correspond to published KFRE variants, with the exception of the 5-lab set, which does not include bicarbonate due to limited availability.

For each laboratory configuration, we trained two XGBoost Survival Embeddings[19] models:

1. **CSARK,** a locally refit static model including demographic features and latest laboratory values only, and
2. **CLARK**, which incorporates trajectory-derived features in addition to latest-value features.

KFRE models were evaluated in the eGFR-only and eGFR-UACR settings. Comparisons between KFRE and CSARK established CSARK as a locally calibrated static baseline. Primary comparisons focused on matched laboratory-set and time-horizon pairs of CSARK and CLARK to isolate the incremental value of longitudinal features. An overview of model comparisons can be found in Figure 2d.

*Data Splitting and Stratification*

Participants were randomly divided into 70% training, 15% validation, and 15% test sets. All splits were stratified by KRT event timing (within 1 year, 1–2 years, 2–5 years, >5 years) and censoring patterns (death or loss to follow-up within the same intervals). The 15% test set was held out for final evaluation and was not used during feature selection or hyperparameter tuning. For each time horizon (2- and 5-year KRT), performance metrics were computed using individuals with known outcomes at that horizon (i.e., excluding those censored prior to the time point).

*Handling Missing and Skewed Data*

For all participants for which latest UACR was unavailable, we created a filled version for use in KFRE using the median value among those in the development set. We also evaluated skew in the development set using the Pandas skew method and transformed features with skew > 1 into log space for use in all models

*Feature Selection*

Feature selection for CLARK models was performed using the 70% training set through greedy additive approach. Highly intercorrelated trajectory-based features were first grouped using hierarchical clustering. Beginning with demographic features (age and sex), candidate groups of trajectory-derived features were evaluated sequentially. At each step, groups were ranked according to improvement in average precision lift, defined as the ratio of model precision to outcome prevalence using internal cross-validation within the training set, and the highest ranked feature group was added. Features were selected optimizing for 2-year performance and 5-year performance separately.

After constructing candidate feature sequences for both time horizons, the final feature set for each laboratory configuration was selected based on improvement in average precision evaluated in the held-out 15% validation set.

*Hyperparameter Tuning*

After feature selection, hyperparameters were optimized using the training and validation sets via HyperOpt, targeting average precision across the two KRT time horizons. Final models were retrained once using the combined training and validation sets.

*Evaluation*

Final model performance was evaluated on the independent 15% test set. Discrimination, assessed via average precision (summarizing how precision varies across all classification thresholds) and c-statistic, and calibration, assessed using the Brier score, were computed for each laboratory configuration and prediction horizon. Confidence intervals were estimated via stratified bootstrapping of the test set (N=100). The same bootstrap resamples were used to compute performance metrics and pairwise differences in average precision between CLARK and CSARK.

Precision-recall curves were generated for all models. Net reclassification improvement was evaluated at clinically relevant KDIGO risk thresholds (≥10% and ≥40% 2-year risk; ≥3% and ≥5% 5-year risk), with $NRI_{KRT}$ defined as the net change in correct classification among individuals who progressed to KRT and $NRI_{No\ KRT}$ defined analogously among individuals who did not. In addition, reclassification was assessed under quantile-based intervention strategies, where interventions were applied to the top k% of predicted risk, using horizon-specific quantiles (0.5, 1, 2, and 3% for 2-year risk and 2, 2.5, 3, 5, and 10% for 5-year risk).

*Feature enrichment analysis*

To characterize the contribution of trajectory-based features to model performance, we performed a post hoc feature enrichment analysis. Following the sequences of feature groups derived during feature selection, we tuned and evaluated average precision for the first three steps of models. These groups included features capturing recent kidney function (e.g., recent averages and trajectory-based estimates at index), baseline kidney function (e.g., long-term and pre-changepoint summaries), and variability in kidney function over time.

At each step, model performance was evaluated on the held-out test set using the same bootstrap resamples described above. This analysis was not used for model training or feature selection, but rather to quantify the incremental contribution of distinct groups of trajectory-based features to overall predictive performance.

**Figure 1: Participant Filtering**

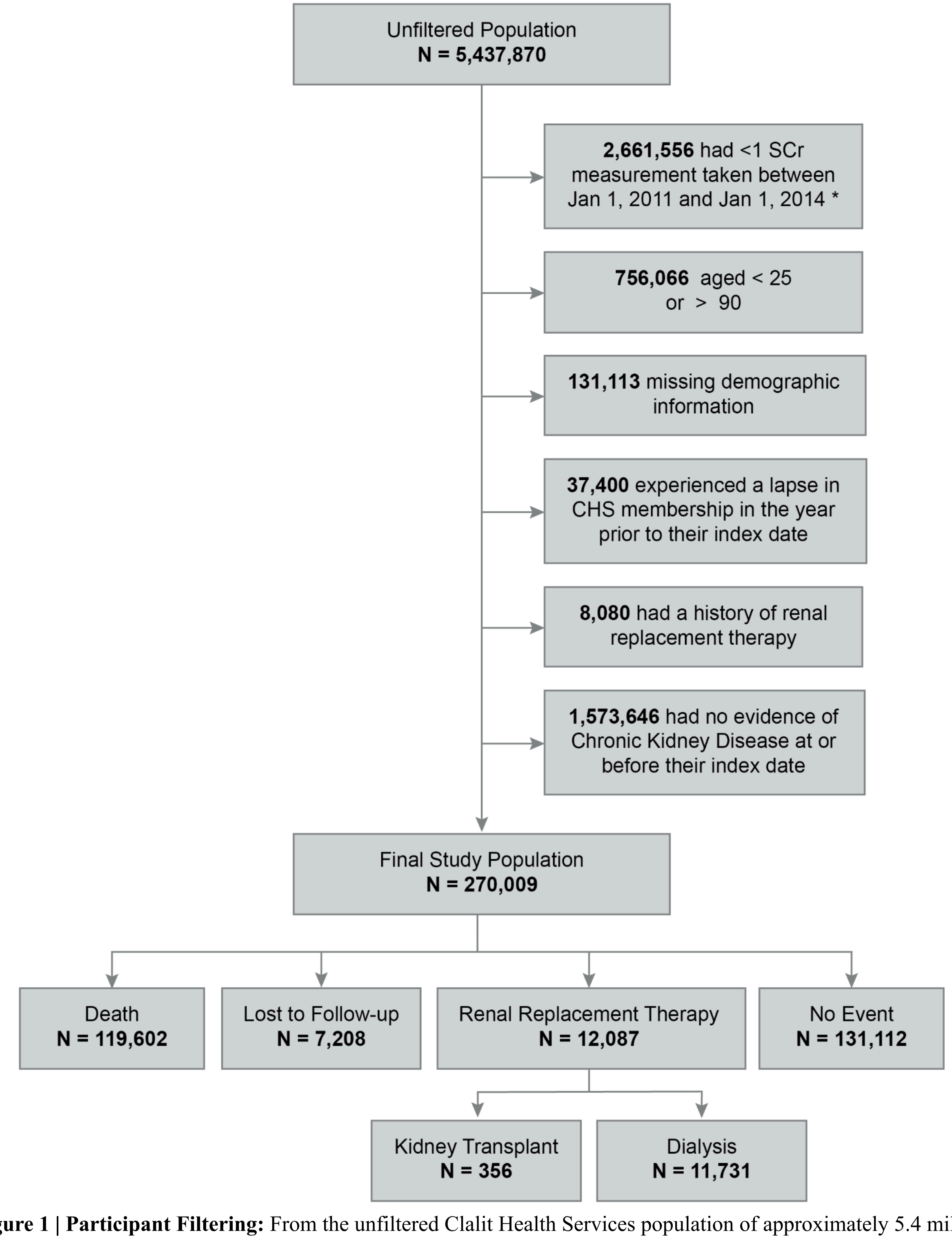

**Figure 1 | Participant Filtering:** From the unfiltered Clalit Health Services population of approximately 5.4 million individuals, we applied inclusion and exclusion criteria to obtain a final cohort of 270,009 adults with chronic kidney disease (CKD). Major exclusion criteria included age outside the range of 25–90, insufficient serum creatinine measurements, missing demographic data, and absence of CKD evidence at index. Outcomes included death, loss to follow-up, initiation of kidney replacement therapy (dialysis or transplant), or no event

## Table 1: Participant Demographics

| | Study Population | KRT | No KRT |
|---|---|---|---|
| Total | 270,009 (100.0%) | 12,087 (100.0%) | 257,922 (100.0%) |
| Female | 144,597 (53.6%) | 4,778 (39.5%) | 139,819 (54.2%) |
| **AGE AT INDEX DATE** | | | |
| Age 25–44 | 12,503 (4.6%) | 647 (5.4%) | 11,856 (4.6%) |
| Age 45–64 | 76,889 (28.5%) | 4,366 (36.1%) | 72,523 (28.1%) |
| Age 65–74 | 68,828 (25.5%) | 3,905 (32.3%) | 64,923 (25.2%) |
| Age 75+ | 111,789 (41.4%) | 3,169 (26.2%) | 108,620 (42.1%) |
| **BIRTH COUNTRY** | | | |
| Israel | 103,957 (38.5%) | 5,524 (45.7%) | 98,433 (38.2%) |
| Central/Eastern Europe & Central Asia | 39,406 (14.6%) | 1,212 (10.0%) | 38,194 (14.8%) |
| North Africa & Middle East | 65,859 (24.4%) | 3,028 (25.1%) | 62,831 (24.4%) |
| Sub Saharan Africa | 3,977 (1.5%) | 169 (1.4%) | 3,808 (1.5%) |
| Other | 56,810 (21.0%) | 2,154 (17.8%) | 54,656 (21.2%) |
| **SOCIOECONOMIC STATUS** | | | |
| Low SES | 81,060 (30.0%) | 3,878 (32.1%) | 77,182 (29.9%) |
| Middle SES | 101,202 (37.5%) | 4,667 (38.6%) | 96,535 (37.4%) |
| High SES | 87,747 (32.5%) | 3,542 (29.3%) | 84,205 (32.6%) |
| **CKD STAGE AND ALBUMINURIA STATUS** | | | |
| G1 | 92,004 (34.1%) | 1,527 (12.6%) | 90,477 (35.1%) |
| G2 | 78,664 (29.1%) | 2,143 (17.7%) | 76,521 (29.7%) |
| G3a | 62,775 (23.2%) | 2,307 (19.1%) | 60,468 (23.4%) |
| G3b | 25,955 (9.6%) | 2,759 (22.8%) | 23,196 (9.0%) |
| G4 | 8,918 (3.3%) | 2,488 (20.6%) | 6,430 (2.5%) |
| G5 | 1,693 (0.6%) | 863 (7.1%) | 830 (0.3%) |
| Albuminuria | 220,312 (81.6%) | 10,759 (89.0%) | 209,553 (81.2%) |
| **EVENT STATUS** | | | |
| Kidney Transplant | 356 (0.1%) | 356 (2.9%) | 0 (0.0%) |
| Dialysis | 11,731 (4.3%) | 11,731 (97.1%) | 0 (0.0%) |
| Death | 119,602 (44.3%) | 0 (0.0%) | 119,602 (46.4%) |
| Lost to Follow-up | 7,208 (2.7%) | 0 (0.0%) | 7,208 (2.8%) |
| First KRT by 2yr | 2,695 (1.0%) | 2,695 (22.3%) | 0 (0.0%) |
| First KRT by 5yr | 6,251 (2.3%) | 6,251 (51.7%) | 0 (0.0%) |
| First KRT by end of study | 12,087 (4.5%) | 12,087 (100.0%) | 0 (0.0%) |
| | | **Median [IQR] – Missing, %** | |
| Follow-up time, years | 10.35 [4.66–10.35] – 0.0% | 4.78 [2.26–7.55] – 0.0% | 10.35 [4.93–10.95] – 0.0% |
| Age, years | 71.78 [61.55–80.25] – 0.0% | 67.53 [59.57–75.42] – 0.0% | 72.08 [61.66–80.48] – 0.0% |
| eGFR, mL/min/1.73 $m^2$ | 75.47 [53.92–95.82] – 0.0% | 44.59 [28.16–67.16] – 0.0% | 76.99 [54.91–96.26] – 0.0% |
| UACR, mg/g | 38.00 [14.37–103.97] – 7.0% | 304.10 [60.81–984.21] – 5.9% | 36.38 [13.99–93.00] – 7.1% |
| Albumin, g/dL | 4.20 [4.00–4.40] – 3.0% | 4.08 [3.80–4.30] – 1.7% | 4.20 [4.00–4.40] – 3.0% |
| Calcium, mg/dL | 9.46 [9.14–9.79] – 1.8% | 9.34 [9.00–9.69] – 0.9% | 9.47 [9.15–9.79] – 1.9% |
| Phosphorus, mg/dL | 3.60 [3.20–3.95] – 3.5% | 3.80 [3.37–4.20] – 2.0% | 3.60 [3.20–3.91] – 3.6% |
| Bicarbonate, mEq/L | 25.10 [22.70–28.00] – 84.0% | 24.40 [21.60–27.50] – 74.5% | 25.10 [22.80–28.10] – 84.5% |

**Table 1 | Participant Demographics:** (CKD, chronic kidney disease; eGFR, estimated glomerular filtration rate; UACR, Urinary Albumin-Creatinine Ratio; KRT, kidney replacement therapy) Characteristics of the study population (n = 270,009), stratified by KRT outcome status. The cohort comprised patients across all stages of chronic kidney disease followed for a median of 10.5 years, during which 12,087 (4.5%) initiated KRT. All characteristics aside from event status and follow-up time represent baseline characteristics at the index date, derived from measurements collected during the historical observational period. Laboratory values represent latest measurements before the index date. Event status and follow-up time were derived from the follow-up period. Values are presented as n (%) for categorical variables and median [IQR] for continuous variables; missingness (%) is reported for continuous variables.

# Figure 2: Study Overview

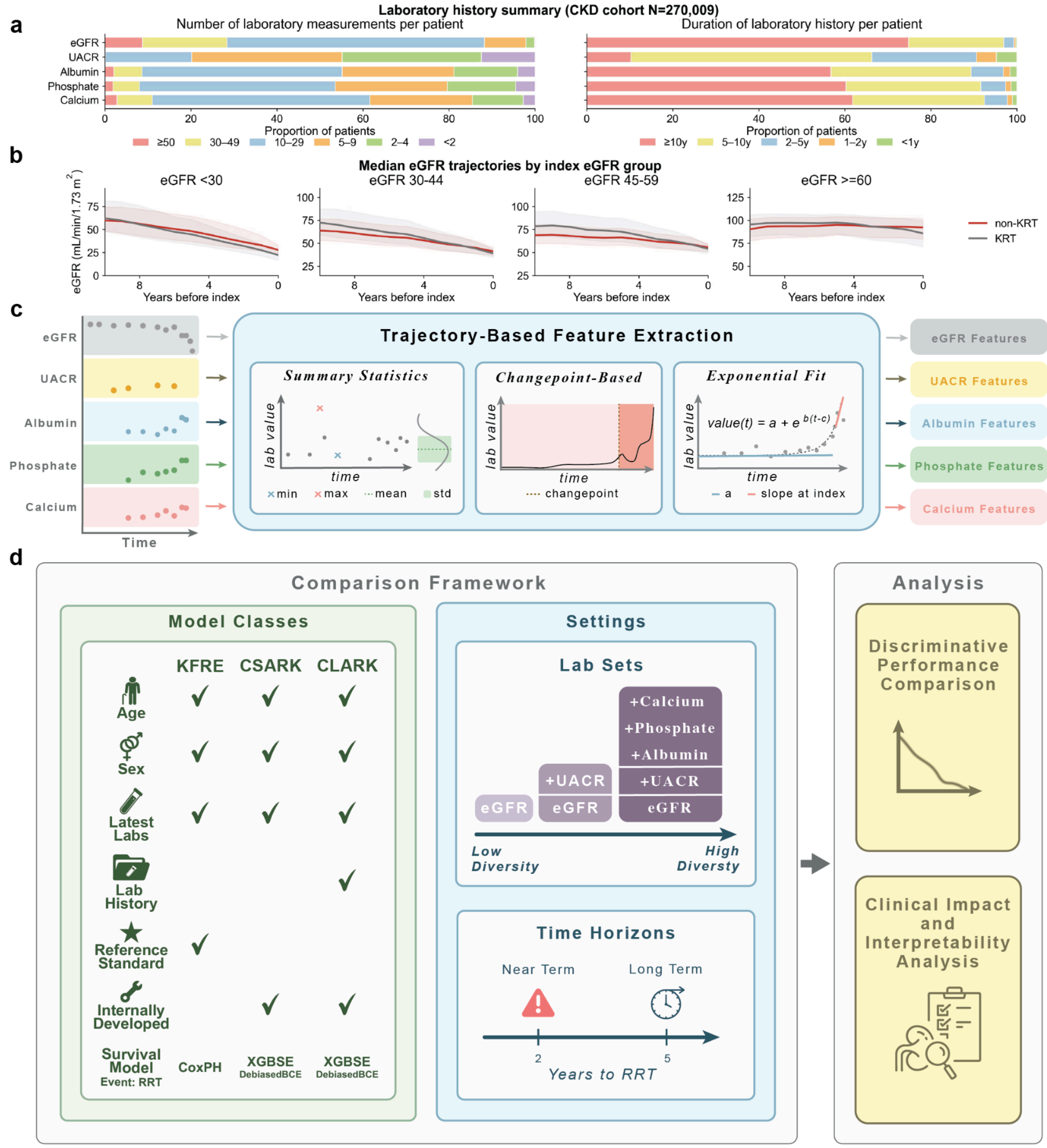

**Figure 2 | Study overview:** (CKD, chronic kidney disease; eGFR, estimated glomerular filtration rate; SCr, serum creatinine; UACR, Urinary Albumin-Creatinine Ratio; KRT, kidney replacement therapy; KFRE, Kidney Failure Risk Equation; CSARK, Clalit Static Assessment of Risk of Kidney Failure; CLARK, Clalit Longitudinal Assessment of Risk of Kidney Failure; CoxPH, Cox proportional hazards; XGBSE, XGBoost Survival Embeddings) (a) Description of dense, extensive longitudinal laboratory data for the study cohort (N=270,009), showing the distribution of the number of measurements per patient (top) and duration of laboratory history (bottom) across each studied lab. Duration is defined as time elapsed between first measurement and index date. (b) Median population-level eGFR trajectories for both progressors and non-progressors aligned by index date. Progressors and non-progressors were stratified by latest eGFR value before index (<30, 30–44, 45–59, ≥60 ml/min/1.73 m$^2$). To obtain median population-level trajectories, eGFR values were binned by year for each participant and the median value for each participant in each year was used. (c) Trajectory-based feature extraction pipeline transforms raw time-series laboratory data into engineered features using three complementary approaches (d) Model comparison framework evaluating guideline-recommended latest-value models (KFRE), locally trained latest-value models (CSARK), and trajectory-based models (CLARK) across laboratory availability and prediction horizons.

## Table 2: Model performance metrics across prediction horizons and laboratory feature sets

| | | KRT within 2 Years | | |
|---|---|---|---|---|
| | | eGFR-only | eGFR + UACR | 5-lab |
| Average Precision | KFRE | 0.520 (0.516, 0.525) | 0.536 (0.531, 0.540) | – |
| | CSARK | 0.516 (0.511, 0.521) | 0.549 (0.545, 0.553) | **0.582 (0.577, 0.586)** |
| | CLARK | **0.541 (0.537, 0.545)** | **0.564 (0.560, 0.568)** | **0.585 (0.581, 0.589)** |
| C-Statistic | KFRE | 0.913 (0.911, 0.914) | 0.918 (0.916, 0.920) | – |
| | CSARK | 0.909 (0.907, 0.911) | 0.917 (0.915, 0.919) | 0.936 (0.934, 0.938) |
| | CLARK | **0.941 (0.939, 0.942)** | **0.937 (0.935, 0.938)** | **0.949 (0.948, 0.951)** |
| Brier | KFRE | 0.0098 (0.0098, 0.0098) | 0.0088 (0.0088, 0.0088) | – |
| | CSARK | 0.0072 (0.0072, 0.0073) | 0.0070 (0.0069, 0.0070) | **0.0067 (0.0066, 0.0067)** |
| | CLARK | **0.0071 (0.0070, 0.0071)** | **0.0068 (0.0068, 0.0069)** | **0.0066 (0.0066, 0.0067)** |

| | | KRT within 5 Years | | |
|---|---|---|---|---|
| | | eGFR-only | eGFR + UACR | 5-lab |
| Average Precision | KFRE | 0.536 (0.533, 0.539) | 0.580 (0.577, 0.583) | – |
| | CSARK | 0.533 (0.530, 0.536) | 0.594 (0.591, 0.597) | 0.617 (0.614, 0.620) |
| | CLARK | **0.567 (0.564, 0.570)** | **0.615 (0.612, 0.618)** | **0.629 (0.626, 0.632)** |
| C-Statistic | KFRE | 0.879 (0.877, 0.880) | 0.898 (0.897, 0.899) | – |
| | CSARK | 0.880 (0.879, 0.882) | 0.910 (0.909, 0.911) | 0.925 (0.924, 0.926) |
| | CLARK | **0.916 (0.915, 0.917)** | **0.925 (0.924, 0.927)** | **0.934 (0.933, 0.935)** |
| Brier | KFRE | 0.0239 (0.0239, 0.0240) | 0.0216 (0.0216, 0.0217) | – |
| | CSARK | 0.0194 (0.0193, 0.0194) | 0.0178 (0.0178, 0.0179) | 0.0174 (0.0173, 0.0175) |
| | CLARK | **0.0187 (0.0186, 0.0187)** | **0.0174 (0.0173, 0.0175)** | **0.0171 (0.0170, 0.0172)** |

**Table 2 | Model performance metrics across prediction horizons and laboratory feature sets:** (AP, Average Precision**;** eGFR, estimated glomerular filtration rate; UACR, Urinary Albumin-Creatinine Ratio; KRT, kidney replacement therapy; KFRE, Kidney Failure Risk Equation; CSARK, Clalit Static Assessment of Risk of Kidney Failure; CLARK, Clalit Longitudinal Assessment of Risk of Kidney Failure; CI, confidence interval) Comparison of AP, C-statistic, and Brier score on the held-out test set for trajectory-based models (CLARK), locally trained latest-value models (CSARK), and established kidney failure risk equations (KFRE) at two prediction horizons: 2-year (top) and 5-year (bottom) risk of KRT. Models are evaluated across three laboratory input configurations: eGFR-only (age, sex, eGFR), eGFR+UACR (age, sex, eGFR, UACR), and 5-lab (age, sex, eGFR, UACR, serum albumin, serum calcium, serum phosphate). Values represent point estimates with 95% CIs from bootstrapped resampling. Bold values indicate best performance within each outcome-lab set combination.

**Figure 3: Average precision comparison for trajectory-based and latest-value kidney failure risk models.**

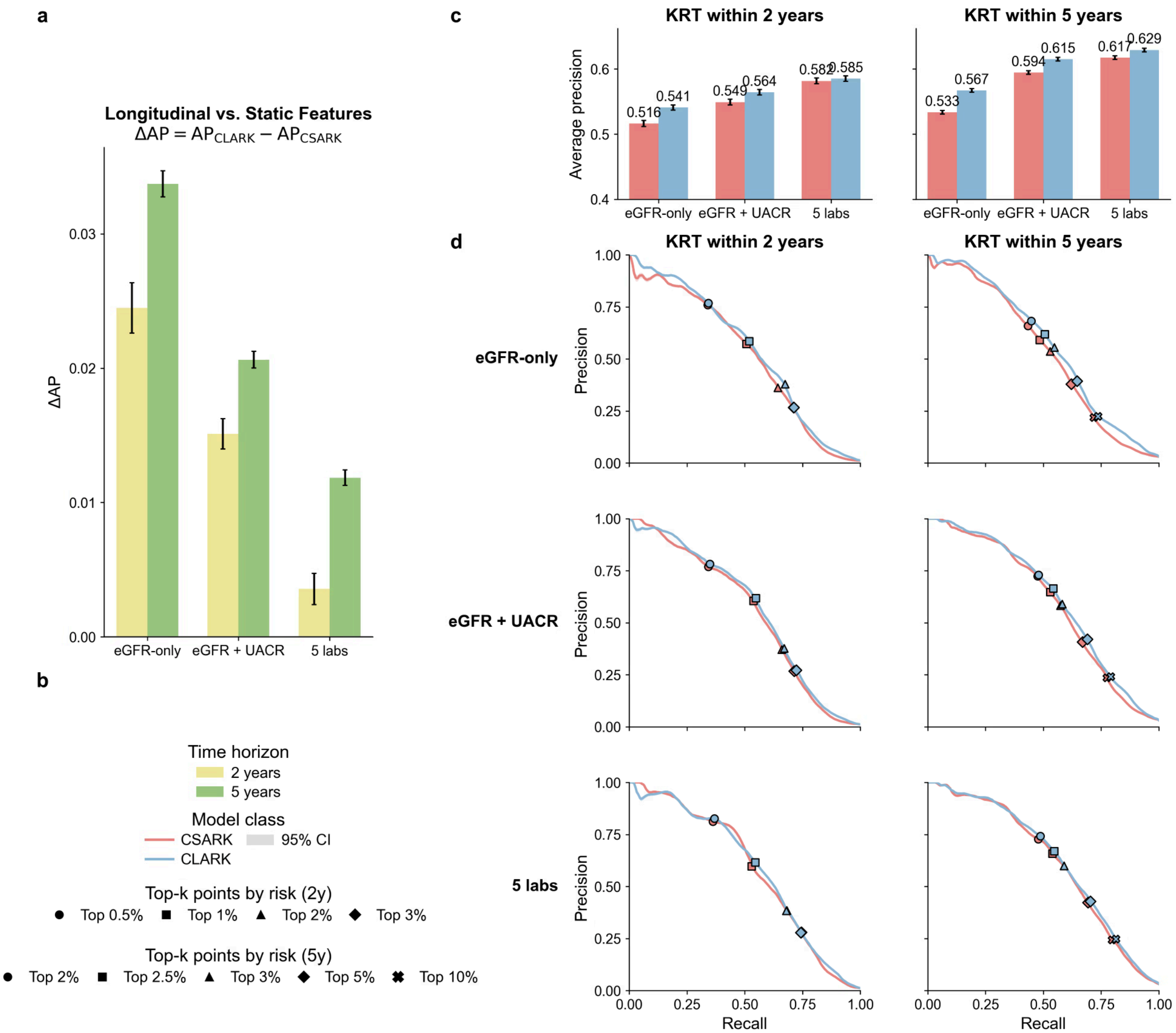

**Figure 3 | Average precision comparison for trajectory-based and latest-value kidney failure risk models:** (AP, Average Precision**;** eGFR, estimated glomerular filtration rate; UACR, Urinary Albumin-Creatinine Ratio; KRT, kidney replacement therapy; KFRE, Kidney Failure Risk Equation; CSARK, Clalit Static Assessment of Risk of Kidney Failure; CLARK, Clalit Longitudinal Assessment of Risk of Kidney Failure; CI, confidence interval) (a) Differences in AP summarizing performance gains from trajectory-based modeling (CLARK - CSARK). Paired differences match lab set and time horizon. Error bars indicate 95% CIs. (b) Legend elements are consolidated across panels: time horizon corresponds to (a); model class and 95% CIs correspond to (c-d); and quantile-based operating points correspond to (d). (c) Mean AP for prediction of KRT within 2 and 5 years across laboratory input sets, comparing locally trained latest-value models (CSARK) and trajectory-based models (CLARK). Error bars show 95% confidence intervals from bootstrapped resampling. Laboratory sets include eGFR-only, eGFR+UACR, and a five-laboratory panel (eGFR, UACR, serum albumin, serum calcium, serum phosphate). (d) Precision-recall curves for eGFR-only, eGFR+UACR, and the five-laboratory panel laboratory input sets (rows) and 2- and 5-year prediction horizons (columns). Shaded regions indicate 95% CIs. Top-k risk threshold points illustrate performance at quantile-based operating points for high-risk patient identification. Horizon-specific quantile thresholds were used to account for differences in outcome prevalence and urgency (0.5–3% for 2-year risk and 2–10% for 5-year risk).

## Figure 4: Clinical impact of risk-based interventions and sources of model improvement

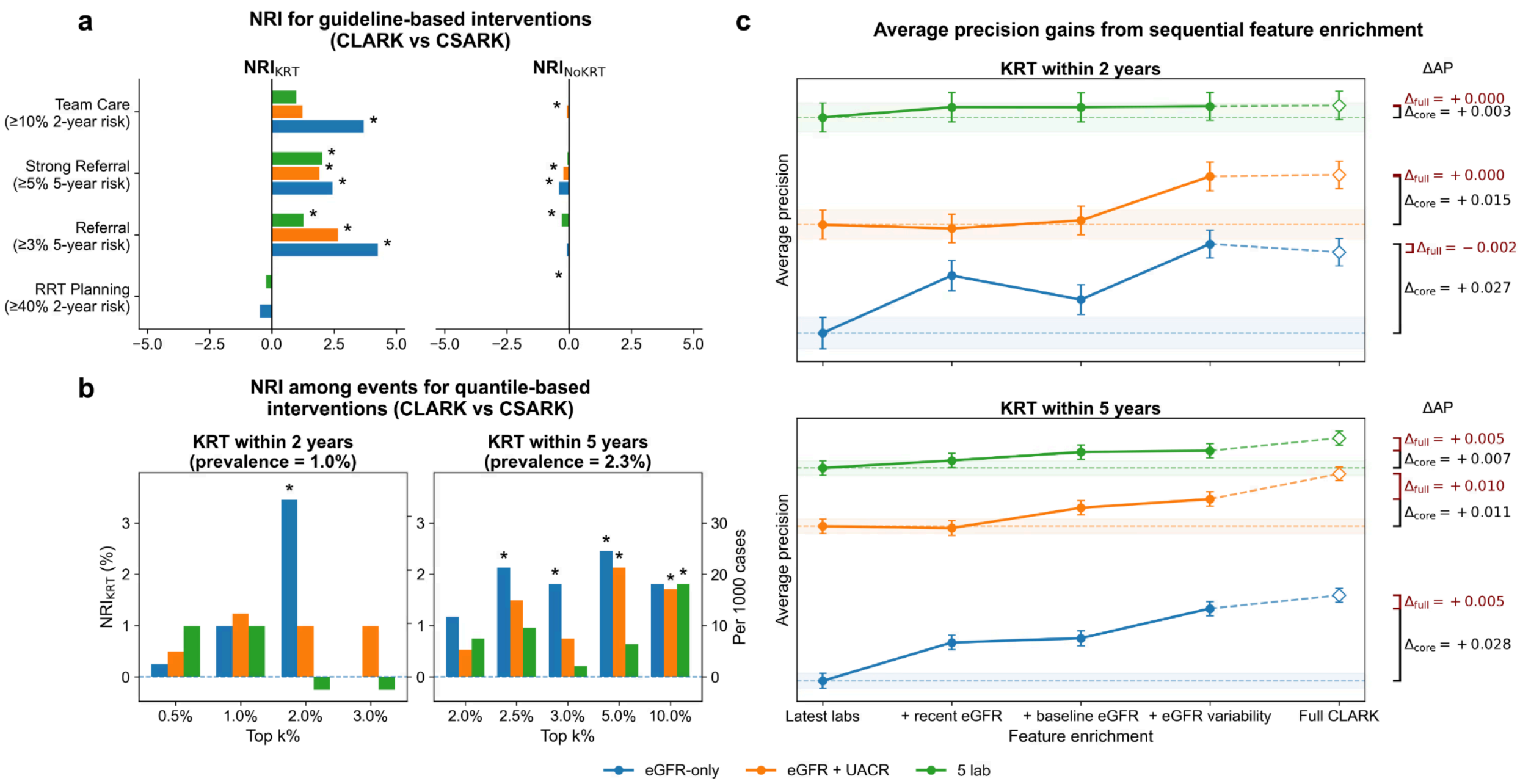

**Figure 4 | Clinical impact of risk-based interventions and sources of model improvement:** (NRI, net reclassification improvement**;** eGFR, estimated glomerular filtration rate; UACR, Urinary Albumin-Creatinine Ratio; KRT, kidney replacement therapy; CLARK, Clalit Longitudinal Assessment of Risk of Kidney Failure; CSARK, Clalit Static Assessment of Risk of Kidney Failure; AP, average precision; CI, confidence interval) (a) NRI comparing CLARK to CSARK across four clinically relevant KDIGO-aligned intervention thresholds: nephrology referral (≥3% 5-year risk), strong referral (≥5% 5-year risk), team care (≥10% 2-year risk), and KRT planning (≥40% 2-year risk). $NRI_{KRT}$ (left) reflects improved identification of individuals who progressed to KRT, while $NRI_{No\ KRT}$ (right) reflects changes among individuals who did not. (b) $NRI_{KRT}$ comparing CLARK to CSARK for quantile-based intervention strategies, where interventions are applied to the top k% of predicted risk. Prevalence values reflect outcome frequency, providing context for intervention scale. Horizon-specific quantile thresholds were used to account for differences in outcome prevalence and urgency (0.5–3% for 2-year risk and 2–10% for 5-year risk). (c) Decomposition of model performance gains from select trajectory-based features across sequential feature enrichment steps and laboratory settings (eGFR-only, eGFR+UACR, and 5-lab). Features were incorporated in groups determined via hierarchical clustering: recent eGFR (3-year pre-index mean, trajectory-based predicted eGFR at index, and post-changepoint mean), baseline eGFR (all-time and pre-changepoint means, and exponential baseline level), and eGFR variability (3-year pre-index standard deviation, absolute and relative to all-time variability). Error bars denote 95% confidence intervals, and ΔAP annotations quantify contributions from core trajectory features and from additional trajectory-based features in the full CLARK models.